\documentclass[twocolumn,showpacs,aps,floatfix,superscriptaddress]{revtex4}
\usepackage{amsmath,amssymb,eucal,graphicx,bm}
\begin{document}
\title{Scaling Exponent for Incremental Records}
\author{P.~W.~Miller} 
\affiliation{Department of Physics, Yale University, New Haven,
  Connecticut 06511 USA}
\affiliation{Theoretical Division and Center for Nonlinear Studies,
  Los Alamos National Laboratory, Los Alamos, New Mexico 87545 USA}
\author{E.~Ben-Naim}
\affiliation{Theoretical Division and Center for Nonlinear Studies,
Los Alamos National Laboratory, Los Alamos, New Mexico 87545 USA}
\begin{abstract}
  We investigate records in a growing sequence of identical and
  independently distributed random variables.  The record equals the
  largest value in the sequence, and our focus is on the increment,
  defined as the difference between two successive records.  We
  investigate sequences in which all increments decrease
  monotonically, and find that the fraction $I_N$ of sequences that
  exhibit this property decays algebraically with sequence length $N$,
  namely $I_N \sim N^{-\nu}$ as $N \rightarrow \infty$. We analyze the
  case where the random variables are drawn from a uniform
  distribution with compact support, and obtain the exponent $\nu =
  0.317621\ldots$ using analytic methods.  We also study the record
  distribution and the increment distribution. Whereas the former is a
  narrow distribution with an exponential tail, the latter is broad
  and has a power-law tail characterized by the exponent
  $\nu$. Empirical analysis of records in the sequence of waiting
  times between successive earthquakes is consistent with the
  theoretical results.
\end{abstract} 
\maketitle

\section{Introduction}

Records, the largest observed values in a sequence of data points
\cite{wf,rse,eig}, have long been a topic of interest --- anybody who
has watched the Olympics will appreciate that much of the suspense
lies in seeing if a new world record will be set.  Records are a
useful tool for characterization of complex systems \cite{jk,gw}, and the
study of records and their statistical properties has proved valuable
in a wide swathe of disciplines, ranging from climate science
\cite{rp,ekbhs,nmt,wzxw} and hydrology \cite{vzm} to economics
\cite{ekm,syn,bp}.

Recent investigations concerning the statistical mechanics of records
\cite{fwk,wbk,ss,wms} reveal rich and interesting phenomenology associated
with the effects of round-off errors on record statistics
\cite{wvrk,ekmb} as well as first-passage behavior \cite{sr,bms} of
record sequences \cite{bk}.  Here, we introduce a first-passage
characteristic that probes how records improve with time, and
demonstrate its usefulness for analysis of empirical data.

Tracking an observable over time, one sees the generation of a
sequence of records, each new record improving upon the previous one
by some finite amount.  Intuitively, we expect these improvements to
diminish with each new record: the larger the current record is, the
less likely it is that the next record improves upon it by a large
amount. In this study we ask: how likely we are to see only shrinking
improvements over previous records?

\begin{figure}[t]
\vspace{.4in}
\includegraphics[width=0.4\textwidth]{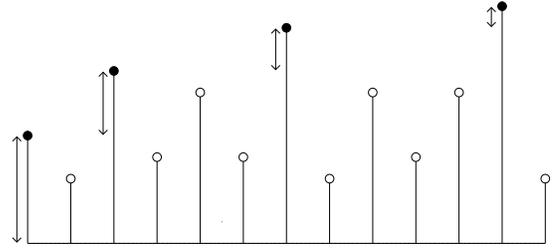}
\caption{Illustration of a sequence with monotonically diminishing
  increments between records. Records are indicated by filled circles,
  and the rest of the variables, by open circles.  The vertical arrows
  to the left of a record indicate 
the magnitudes of the respective increment.}
\label{fig-ill}
\end{figure}

Consider an evolving sequence of $N$ identical and independently
distributed random variables.  The first element of this sequence is,
by definition, a record. Each subsequent element is a record if it is
greater than the previous record.  Moreover, a new record improves
upon the existing one by some increment, defined as the difference
between the two records. We study the probability $I_N$ that all
increments decrease monotonically (see Fig.~\ref{fig-ill}), and our
main result is the scaling law
\begin{equation}
\label{FN-decay}
I_N \sim N^{-\nu}
\end{equation}
which holds in the large-$N$ limit.  When the random variables are
drawn from a compact, uniform distribution, we obtain analytically the
exponent
\begin{equation}
\label{nu}
\nu = 0.317621\ldots .
\end{equation}

Our definition of incremental records involves the current record and
the previous record, or alternatively, the current record and the
current increment.  Consequently, the theoretical analysis requires
the joint distribution of these two variables.  Interestingly, the
record and the increment become uncorrelated in the large-$N$ limit,
such that the joint distribution is a product of the record
distribution and the increment distribution.  The former distribution
is exponential, but the latter is broad and has power-law tail. 

The rest of this paper is organized as follows. In section II, we 
examine the record distribution, and show that it is the same
regardless of whether one considers the set of all records sequences
or just the subset of incremental ones. Next, we study the increment
distribution which in turn requires the full joint distribution of
records and increments (section III). We obtain the increment
distribution in the scaling limit, and derive the scaling exponent
$\nu$ as a byproduct. In Section IV, we measure the fraction of
incremental records for the sequence of waiting times between large
earthquakes, and observe good agreement with the theoretical
results. We conclude in Section V.

\section{Record Distribution}

Let us consider a sequence of uncorrelated random variables,
\begin{equation}
\label{sequence}
\{X_1,X_2,\ldots,X_N\}.
\end{equation}
Each variable $X_i\geq 0$ is independently drawn from the probability
distribution function $\rho(X)$, with the normalization $\int_0^\infty
dX\rho(X)=1$. In many applications, including the earthquake example
discussed in section IV, new variables are constantly added to the
dataset. Hence, we may view the sequence \eqref{sequence} as evolving
with the variable $N$ playing the role of time \cite{krb}.

The record $x_i$ equals the largest variable in a sub-sequence of
length $i$.  A newly added variable sets a record if it is larger than
the previous record,
\begin{equation}
\label{record}
x_{N+1} =
\begin{cases}
x_N       &  \quad x_N\geq X_{N+1},\\
X_{N+1}   &  \quad X_{N+1}>x_N,
\end{cases}
\end{equation}
for $N\geq 1$ with $x_0=0$.  The first variable necessarily sets a
record, \hbox{$x_1=X_1$}, and by definition, the records increase
monotonically, \hbox{$x_{N+1}\geq x_N$}.

We focus on the simplest case of a uniform distribution with support
in a finite interval, taken without loss of generality as the unit
interval,
\begin{equation}
\label{uniform}
\rho(X) =
\begin{cases}
1   &  \quad 0\leq X\leq 1,\\
0   &  \quad  1<X.
\end{cases}
\end{equation}
This distribution is relevant for waiting times between successive
earthquakes (see section IV). 

The cumulative record distribution $R_N(x)$ equals the probability
that the record $x_N$ is smaller than $x$. This distribution is simply 
\begin{equation}
\label{RNx}
R_N(x) = x^N.
\end{equation}
Indeed, by substituting \eqref{uniform} into $R_1(x)=\int_0^x
dX\,\rho(X) $ we have $R_1(x)=x$, and further, for the $N$th record to
be smaller than $x$, all $N$ variables must be smaller than $x$.

Since the variables are identical and independently distributed, every
one of the $N$ variables is equally likely to be the largest. Hence,
the probability that the $N$th variable sets a record equals
$\frac{1}{N}$, and consequently, the average number of records equals
the harmonic number
\hbox{$1+\frac{1}{2}+\frac{1}{3}+\cdots+\frac{1}{N}$}. The slow
logarithmic growth of this sum demonstrates that when $N$ becomes
large, new records are few and far between \cite{gkp,bhi,gl}.

Further, we expect that the improvements made with each new record
diminish with time. The increment, defined as the difference between
the new record and the old record, quantifies such
improvements. Together with the current record $x_N$ we also track the
{\em current increment} $y_N$ defined by
\begin{equation}
\label{increment}
y_{N+1}=
\begin{cases}
y_N           &  \quad x_{N+1}=x_N,\\
x_{N+1}-x_N   &  \quad x_{N+1}>x_N,
\end{cases}
\end{equation}
for $N\geq 1$.
Again, $y_1=x_1$ since the first variable is a record. 

In this study, we restrict our attention to the subset of sequences
where all increments decrease monotonically,
\begin{equation}
\label{incremental}
y_1\geq y_2 \geq y_3 \geq \cdots \geq y_N.
\end{equation}
If this inequality holds, the sequence of records
$\{X_1,X_2,\ldots,X_N\}$ is said to be {\em incremental}.  We are
interested in the probability $I_N$ that a record sequence of length
$N$ is incremental. Essentially, we require that the quantity
$y_{N}-y_{N+1}$ remains non-negative. In this sense, the condition
\eqref{incremental} defines a first-passage process, and the quantity
$I_N$ is analogous to a survival probability \cite{sr}.

As a preliminary step, we analyze the distribution of incremental
records. Specifically, we define the cumulative density $F_N(x)$ as
the fraction of sequences that: (i) have incremental records, namely,
satisfy the condition \eqref{incremental}, and (ii) have a record
smaller than $x$, that is, $x_n<x$.  Of course, $I_N\equiv F_N(x=1)$.

\begin{figure}[t]
\includegraphics[width=0.3\textwidth]{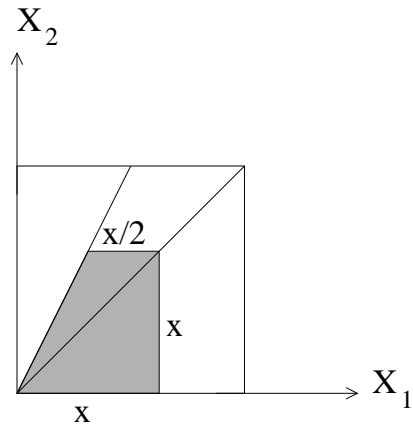}
\caption{Geometric representation of incremental sequences of length
two. The shaded area represents sequences $\{X_1,X_2\}$ with
incremental records and current record $x_2<x$.}
\label{fig-trapezoid}
\end{figure}

When $N=1$, we have $F_1(x)=R_1(x)$ and hence, $I_1=1$. When $N=2$,
the two-variable sequence $\{X_1,X_2\}$ corresponds to a point inside
the unit square (figure \ref{fig-trapezoid}). If $X_1>X_2$, the record
sequence is necessarily incremental. Otherwise, the sequence is
incremental if and only if $X_2-X_1\leq X_1$. Accordingly, the lines
$X_2=X_1$ and $X_2=2X_1$ divide the unit square into three triangles,
in two of which the condition \eqref{incremental} holds.  To obtain
the probability $F_2(x)$ that an incremental sequence has record
$x_2\leq x$, we overlay a square of area $x^2$ onto the unit square.
The overlap between this square and the two relevant triangles is a
trapezoid with height $x$, and bases $x$ and $x/2$ (figure
\ref{fig-trapezoid}).  The area of this trapezoid gives the cumulative
record density
\begin{equation}
\label{F2}
F_2(x)=\frac{3}{4}\,x^2, 
\end{equation}
and hence, $I_2=\tfrac{3}{4}$.

\begin{figure}[t]
\includegraphics[width=0.475\textwidth]{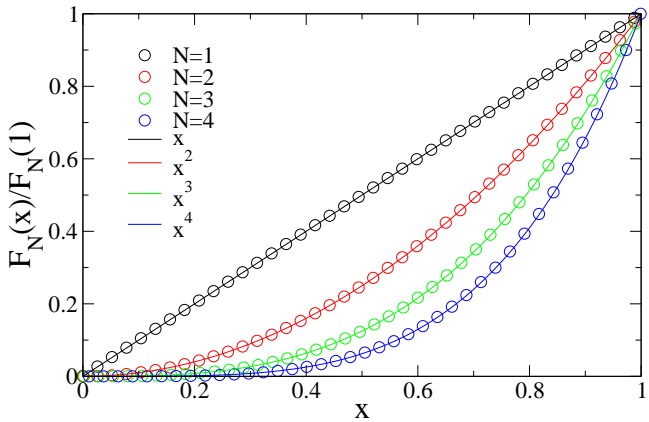}
\caption{The normalized cumulative distribution $F_N(x)/F_N(1)$ for
  $N\leq 4$. The circles represents results of Monte Carlo simulations
  and the lines show the corresponding monomials $x^N$.}
\label{fig-FNX}
\end{figure}

This illuminating example suggests that generally, 
\begin{equation}
\label{FNX}
F_N(x)=I_N\,x^N.
\end{equation}
Indeed, when $N=3$, in addition to the planes $X_2=X_1$, $X_2=2X_1$,
and similarly, $X_3=X_1$ and $X_3=2X_1$, there is a fifth plane
$X_3-X_2=X_2-X_1$ that corresponds to three distinct records. These
planes specify a polyhedron with volume $F_3(x)=\tfrac{47}{72}\,x^3$,
embedded inside a cube of volume $x^3$; The condition
\eqref{incremental} holds inside this polyhedron.  In general, the
hyperplanes defined by \eqref{incremental} are invariant under the
transformation $X_i\to a X_i$.  Consequently, the cumulative
distribution equals the volume of an $N$-dimensional polyhedron,
embedded in an $N$-dimensional hypercube of volume $x^N$.  This
geometric argument and the normalization $F_N(1)=I_N$ support equation
\eqref{FNX} (numerical confirmation is shown in figure \ref{fig-FNX}).
Remarkably, the normalized distribution function $F_N(x)/F_N(1)$ is
identical to $R_N(x)$ given in \eqref{RNx}. Hence, records in
incremental sequences and records in all sequences are characterized
by {\em identical} distribution functions.

\begin{figure}[t]
\includegraphics[width=0.475\textwidth]{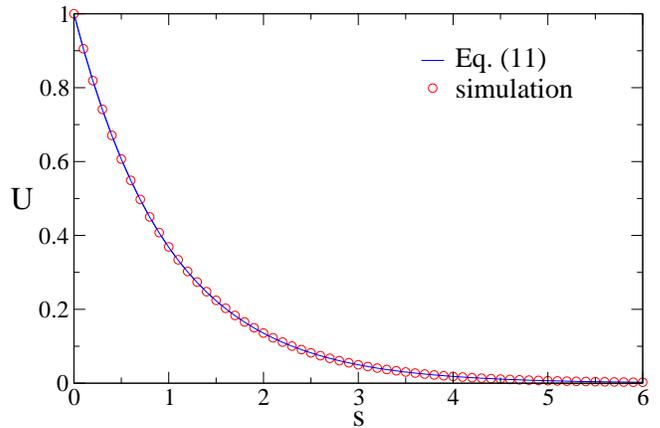}
\caption{The scaling function governing the records.  The circles
  represent results of Monte Carlo simulations for the fraction $U(s)$
  of incremental sequences with scaled record $(1-x)N$ that is larger
  than $s$. Also shown for reference is the prediction $U(s)=\exp(-s)$
  that follows from Eq.~\eqref{FN-scaling}.}
\label{fig-FN-scaling}
\end{figure}

Our main focus is the asymptotic behavior in the limit $N\to\infty$
where the cumulative density \eqref{FNX} follows the scaling form
(figure \ref{fig-FN-scaling})
\begin{equation}
\label{FN-scaling}
F_N(x)\simeq I_N\, e^{-s}, \qquad s=(1-x)N.
\end{equation}
To obtain this form, we simply rewrite \hbox{$x^N=[1-(1-x)]^N$} and
consider the limits $N\to\infty$ and $x\to 1$ with the scaling
variable $s=(1-x)N$ being finite.  Thus, the term $x^N$ implies that
the record distribution is exponential \cite{ft,eig1}.

\section{Increment Distribution}

As evident from the definitions \eqref{record} and \eqref{increment},
the increment is coupled to the record. Hence, further analysis
requires the joint record-increment distribution. We study the
probability density $S_N(x,y)$, defined such that $S_N(x,y)\,dx\,dy$
is the probability that the record $x_N$ lies in the infinitesimal
range $(x,x+dx)$ and similarly, the increment $y_N$ is in the range
$(y,y+dx)$. Since $y<x$, the fraction of incremental sequences is the
following integral of the joint density,
\begin{equation}
\label{SN-int}
I_N=\int_0^1 dx \int_0^x dy\, S_N(x,y).
\end{equation}

The probability density satisfies the recursion equation 
\begin{equation}
\label{SN-rec}
S_{N+1}(x,y)=x\,S_N(x,y)+\int_y^{x - y} dy'\,S_N(x-y,y')
\end{equation}
for $N> 1$, with $S_1(x,y)=\delta(x-y)$ as the first element is
necessarily a record. The first term on the right-hand side of
\eqref{SN-rec} accounts for cases where the old record holds. The
integral term accounts for situations where the old record, which
necessarily equals $x-y$, is surpassed by a new record; The lower
integration limit enforces the monotonicity condition
\eqref{incremental}.  Starting with $S_1(x,y)=\delta(x-y)$ we find
$S_2(x,y)=x\delta(x-y)+\Theta(x-2y)$ from which $I_2=3/4$ is 
recovered.

To study the asymptotic behavior for large $N$, we convert the
recursion equation \eqref{SN-rec} into the integro-differential
equation 
\begin{equation}
\label{SN-eq}
\frac{\partial S_N(x,y)}{\partial N} = -(1-x)S_N(x,y)+\int_y^{x-y}
dy'\,S_N(x-y,y').
\end{equation}
To obtain this equation, we subtract $S_N(x,y)$ from both sides of
\eqref{SN-rec} and then replace the difference $S_{N+1}-S_N$ with the
derivative $\partial S_N/\partial N$. 

The scaling behavior \eqref{FN-scaling} shows that $1-x\sim
N^{-1}$. Moreover, the integral term in \eqref{SN-eq} implies that, 
similarly, the increment is inversely proportional to sequence length,
$y\sim N^{-1}$. Hence, we anticipate the scaling form
\begin{equation}
\label{SN-scaling}
S_N(x,y) \simeq I_N\,N^2\,\Psi(s,z),\qquad z=y\,N.
\end{equation}
The normalization \eqref{SN-int} sets the prefactor $I_N\,N^2$ and
implies that the scaling function $\Psi(s,z)$ is normalized,
$\int_0^\infty \int_0^\infty ds\,dz\,\Psi(s,z) = 1$. If we integrate
the joint scaling function over the scaled increment, we should
recover the distribution of the scaled record, given in
\eqref{FN-scaling},
\begin{equation}
\label{exponential}
\int_0^\infty dz\, \Psi(s,z)=e^{-s}.
\end{equation}

\begin{figure}[t]
\includegraphics[width=0.475\textwidth]{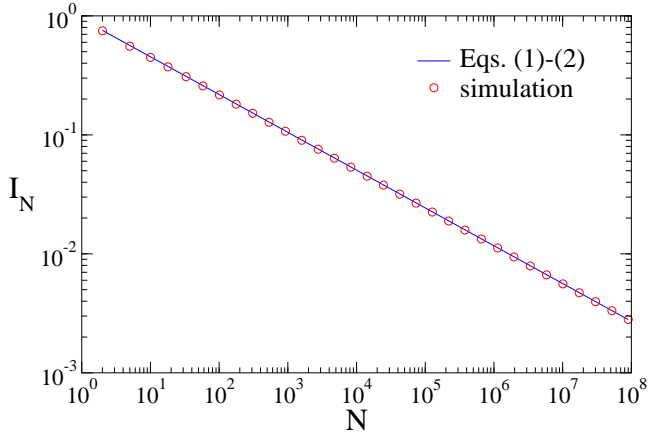}
\caption{The fraction of incremental sequences versus sequence length.
  The quantity $I_N$, measured from Monte Carlo simulations is 
  compared with the theoretical result 
  \eqref{FN-decay}-\eqref{nu}. The numerical results are obtained 
  from $10^8$ independent realizations.}
\label{fig-sn}
\end{figure}

Next, we substitute \eqref{SN-scaling} along with $I_N\simeq
A\,N^{-\nu}$ as in \eqref{FN-decay} into the evolution equation
\eqref{SN-eq}, and find that the scaling function satisfies 
\begin{equation}
\label{psi-eq}
\left(\!2\!-\!\nu\!+\!s\!+\!s\frac{\partial}{\partial s} 
\!+\!z\frac{\partial}{\partial z}\!\right)\!\Psi(s,z\!) = \!
\int_z^\infty\!\! dz'\Psi(s+z,z').
\end{equation}
Remarkably, this rather involved integro-differential equation admits
a separation-of-variables solution
\begin{equation}
\label{factor}
\Psi(s,z)=e^{-s}\,\phi(z).
\end{equation}
This form is consistent with Eq.~\eqref{exponential} when the scaling
function $\phi(z)$ that characterizes the distribution of increments,
is normalized, $\int_0^\infty dz\,\phi(z)=1$.  The factorizing form 
\eqref{psi-eq} implies that records and increments become uncorrelated
in the limit $N\to\infty$, an assumption that is supported by the
numerical simulations (see below).

\begin{figure}[t]
\includegraphics[width=0.475\textwidth]{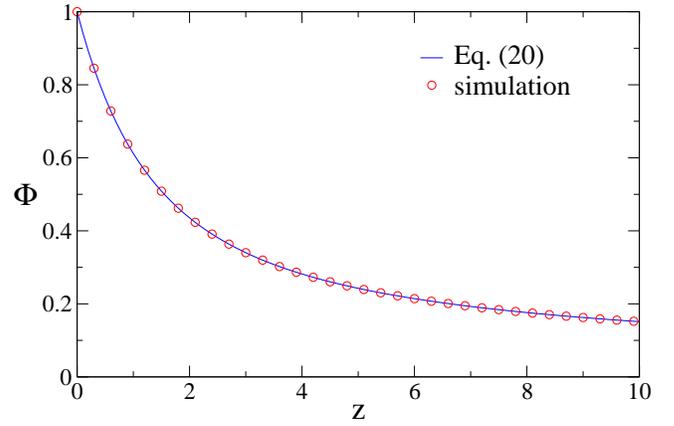}
\caption{The cumulative distribution of increments. Shown is the
probability $\Phi(z)$ that the scaled increment $yN$ is larger than
the scaling variable $z$. The simulation results were obtained by
generating $10^7$ independent increments for a sequence of length
$N=10^6$. The theoretical result represents a numerical solution of
\eqref{Phi-eq} with $\nu$ given by Eq.\eqref{nu}.}
\label{fig-compare}
\end{figure}

By substituting the factorizing form \eqref{factor} into the governing
equation \eqref{psi-eq}, we obtain the integro-differential equation 
\begin{equation}
\label{phi-eq}
z \frac{d\phi(z)}{dz} + (2 - \nu) \phi(z) = e^{-z} \int_z^\infty \phi(z') dz'.
\end{equation}
We can convert this first-order equation for $\phi(z)$ into a second
order ordinary differential equation for its integral,
$\Phi(z)=\int_z^\infty dz'\,\phi(z')$,
\begin{equation}
\label{Phi-eq}
z \frac{d^2\Phi(z)}{dz^2}+(2-\nu)\frac{d\Phi(z)}{dz}+e^{-z}\Phi(z)=0.
\end{equation}
The two boundary conditions are $\Phi(0)=1$ and
\hbox{$\Phi'(0)=-1/(2-\nu)$}. The former follows from the fact that
$\phi(z)$ is normalized, and the second boundary condition follows
from equation \eqref{phi-eq} itself.

Equation \eqref{Phi-eq} has two independent solutions. In the
large-$z$ limit, the rightmost term in Eq.~\eqref{Phi-eq} vanishes,
and consequently, one of these solution approaches a constant,
$\Phi(z) \to C$, while the other decays algebraically,
\begin{equation}
\label{Phi-tail}
\Phi(z) \sim z^{\nu - 1}.
\end{equation}
This former solution is not physical because the scaling function
$\phi(z)$ vanishes in the limit $z\to \infty$.  The exponent $\nu$
plays the role of an ``eigenvalue'' of equation \eqref{Phi-eq}.
Numerically, we integrate \eqref{Phi-eq} with a trial value $\nu_{\rm
  try}$. When $\nu_{try}\neq \nu$, the solution approaches a nonzero
constant $C\equiv C(\nu_{\rm try}) \neq 0$ when $z\to\infty$. Only
when $\nu_{\rm try}$ equals the eigenvalue $\nu$, does this constant
vanish $C(\nu)=0$, and the physical solution \eqref{Phi-tail} is
realized.  We used the Adams method \cite{ba} to perform the numerical
integration and the bisection method \cite{nr} to find the root of the
equation $C(\nu)=0$. The result $\nu=0.31762101\ldots$ quoted in
\eqref{nu} follows.

Equation \eqref{Phi-tail} implies that the increment distribution
decays algebraically for sufficiently large increments. Let $P_N(y)dy$
be the probability that an incremental sequence has latest increment
in the range $(y,y+dy)$. The probability density $P_N(y)$ follows
from the joint density,
\begin{equation}
\label{PN-def}
P_N(y)=\int_y^1 dx\,  S_N(x,y).
\end{equation}
By substituting the scaling behavior \eqref{SN-scaling} with the
factorizing form \eqref{factor} into the integral, we find that the
increment density has the scaling form
\begin{equation}
\label{PN-scaling}
P_N(y)\simeq I_N\,N\,\phi(z).
\end{equation}
The scaling function $\phi(z)$ is simply
$\phi(z)=-\Phi'(z)$. Therefore, using the algebraic tail
\eqref{Phi-tail}, we find \hbox{$\phi(z)\sim z^{\nu-2}$}. Hence, the
increment density decays algebraically,
\begin{equation}
\label{PN-tail}
P_N(y)\sim N^{-1}y^{\nu-2},
\end{equation}
for sufficiently large increments $y\gg N^{-1}$. 

The algebraic decay \eqref{PN-tail} matches our expectation that is
based on a simple heuristic argument. The probability that the first
element is largest equals $N^{-1}$. Since a sequence with only one
record is incremental, we expect that $P_N(1)\sim N^{-1}$. This
behavior agrees with \eqref{PN-tail}. Hence, from the extreme case of
a single record that is nearly maximal, $y\approx 1$, we can derive
the algebraic behavior \eqref{PN-tail} from the scaling behavior
\eqref{PN-scaling} together with \eqref{FN-decay}. We note that the
same argument applies to all record sequences. This heuristic argument,
combined with the fact that $F_N(x=1)\equiv 1$, yields $P_N(y)\sim
N^{-1}y^{-2}$ for all record sequences.

We performed numerical simulations to test the above predictions.  In
each simulation run, the initial record and the initial increment are
respectively set as $x_0=0$ and $y_0=1$. Then at each step a random
number in the range $[0:1]$ is drawn \cite{nr}. The record and the
increment are calculated from \eqref{record} and \eqref{increment}
respectively. This elementary step is iterated as long as the
monotonicity condition $y_{N+1}\leq y_N$ is satisfied, but the run is
aborted when this condition is violated for the {\em first} time. For
example, the quantity $I_N$ is measured as the fraction of such runs
where at least $N+1$ random numbers where generated.

Results of these Monte Carlo simulations confirm the theoretical
prediction. First, as shown in figure \ref{fig-sn}, the fraction of
incremental sequences decays algebraically as in equations
\eqref{FN-decay}-\eqref{nu}. Second, as shown in figure
\ref{fig-compare}, the probability $\Phi(z)$ that the scaled increment
$yN$ is larger than $z$ agrees with the predictions of equation
\eqref{Phi-eq}, with the exponent $\nu$ specified in
\eqref{nu}. Finally, we also confirmed directly that the record and
the increment become uncorrelated, that is, $\langle (1-x)y\rangle
/\langle 1-x\rangle \langle y\rangle\to 1$ when $N\to\infty$.

\begin{figure}[t]
\includegraphics[width=0.475\textwidth]{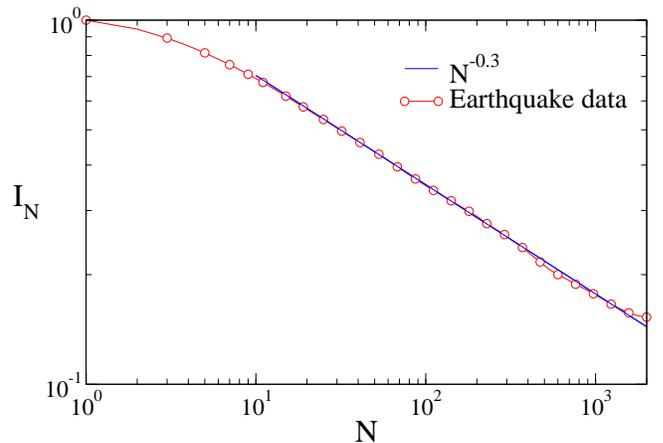}
\caption{The probability $I_N$ as calculated using data for
  earthquakes of magnitude $4$ and above. The empirical results are
  compared with theoretical predictions for the compact, uniform
  distribution.}
\label{fig-earth}
\end{figure}

\section{Empirical Study}

Records are a basic feature of a data set, as for example, the record
high and record low specify the span of the set. It is natural to ask
whether the statistical measures introduced in this study are useful
as a data analysis tool. We believe that the notion of incremental
records is a sensible measure of performance \cite{bk}, especially
since it involves no prior knowledge of how the random variables are
distributed.  Also, determining whether a sequence includes only
incremental records is straightforward. There is a difficulty,
however.  In practice, a very large number of data points is required
to accurately measure the fraction $I_N$ because this survival
probability decays as a power-law.

Here, we analyze inter-event times between successive earthquakes. In
ref.~\cite{bk}, it was demonstrated that the average number of records
in inter-event times is a straightforward test for whether powerful
earthquakes occur randomly in time and follow Poisson statistics
\cite{am,st,bdj}.  Inter-event times are not bounded from above, but
they are bounded from below, by zero. Given that our theoretical
analysis applies to bounded random variables, rather than studying
increments in the record high, we studied decrements in the record
low.  As the transformation $X_N\to 1-X_N$ shows, the behavior
\eqref{FN-decay}-\eqref{nu} also characterizes decrements in record
lows.

Our dataset lists the event times for the $\approx 189,000$
earthquakes with magnitude $M>4$ \cite{gr} that occurred worldwide
during the years $1984-2012$ \cite{data}. From these event times, the
sequence \eqref{sequence} of inter-event times is constructed as a
list of the waiting times between consecutive earthquakes (zero
waiting times are ignored).  Our procedure for determining $I_N$ from
this data set is to treat $N$ consecutive data points as an
independent sequence, and then calculate the fraction of these which
have only incremental records. That is to say, when calculating $I_1$
we consider each individual element of the full data set to be a
sequence of length $1$, when calculating $I_2$ each consecutive pair
of events to be a distinct sequence of length $2$, and so forth.

As Fig.~\ref{fig-earth} shows, we observe scaling behavior which
closely matches the theoretical predictions
\eqref{FN-decay}-\eqref{nu}. The  measured exponent is
remarkably close to the theoretical value
\begin{equation}
\nu_{\rm data}=0.30\pm 0.03. 
\end{equation}
As discussed in the conclusions, the tail of the distribution function
from which the random variables are drawn controls $\nu$
\cite{bk}. Hence, the empirical data combined with the theoretical
results above suggests that there are no singularities (due to
aftershocks) in the distribution of inter-event times, $\lim_{X\to
0}\rho(X)={\rm const}$. Moreover, the empirical results show that the
fraction of incremental records is a sensible quantity in the context
of data analysis, and provide an example where the fraction $I_N$
decays algebraically.

\begin{figure}[t]
\includegraphics[width=0.475\textwidth]{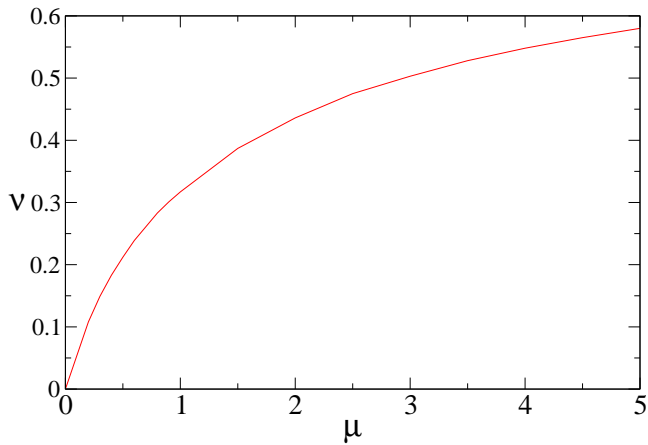}
\caption{The exponent $\nu$ versus the parameter $\mu$ characterizing
  the parent distribution $\rho(X)$ in \eqref{rho}.}
\label{fig-numu}
\end{figure}

\section{Conclusions} 

In conclusion, we studied the probability that all records in a
sequence of random variables are incremental. In our definition, a
record is incremental if it improves upon the previous record by a yet
smaller amount. We found two scaling laws: one for the probability
that a sequence is incremental, and one for the distribution of
increments. A single scaling exponent underlies both of these scaling
laws. Interestingly, even for the simplest case of identical and
independently distributed random variables, drawn from a uniform
distribution, this scaling exponent is nontrivial. Using an empirical
analysis of earthquake data, we demonstrated that statistics of
incremental records are a sensible tool for data analysis.

Our theoretical analysis involved the joint distribution of record
value and increment size. Initially, these two variables are perfectly
correlated, but as the sequence grows, the correlation between these
two variables diminishes, and it eventually vanishes. This key
observation allows us to solve for the full distribution of records
and increments in the scaling limit and obtain as a byproduct the
scaling exponent.

Our study concerns an ensemble of sequences for which improvements in
records diminish steadily with time.  It is useful to compare
statistical properties of records and increments for this restricted
ensemble of sequences with the corresponding behavior for the full
ensemble of sequences \cite{krb}.  Our findings show that the most
basic characteristic, the record, does not distinguish between these
two ensembles, yet, differences in the record are a more sensitive
measure whose distribution does differentiate between the two.

In a recent related study, it was shown that another measure for
performance, the probability that records always remain above their
expected average, is also characterized by a nontrivial scaling
exponent \cite{bk}. Taken together, these two closely related studies
indicate that there is a family of first-passage exponents \cite{bk1}
that characterize extreme statistics.

An interesting challenge is to generalize the above results to
arbitrary distributions, both compact and noncompact. Numerically, we
also studied the general class of compact distribution functions with
algebraic tail
\begin{equation}
\label{rho}
\rho(X)=
\begin{cases}
\mu (1-X)^{\mu-1}   &  \quad 0\leq X\leq 1,\\
0                   &  \quad     1<X,
\end{cases}
\end{equation}
for $0\leq X\leq 1$. The simulations show that the exponent $\nu$
varies continuously with the parameter $\mu$ (figure
\ref{fig-numu}). Qualitatively, this behavior is in line with the
results of several other studies showing that the tail of the
distribution $\rho$ governs scaling laws for extreme statistics
\cite{bk,bkr,kj}. It will be interesting to establish whether the
increment and the record decouple in general, and to obtain the
scaling exponent $\nu$ using analytic methods.

\bigskip \acknowledgements We thank Paul Krapivsky for collaboration
in early stages of this work, Ivan Christov and Joan Gomberg for
useful discussions, and Chunquan Wu for assistance with the earthquake
data. We also acknowledge support from US-DOE through grant
DE-AC52-06NA25396 and the SULI program.


\begin{thebibliography}{99}

\bibitem{wf} 
      W.~Feller, 
      {\it An Introduction to Probability Theory and Its Applications}
      (Wiley, New York, 1968).

\bibitem{rse}
       R.~S.~Ellis, 
       {\it Entropy, Large Deviations, and Statistical Mechanics}
       (Springer, Berlin 2005).

\bibitem{eig}
       E.~I.~Gumbel, 
       {\it Statistics of Extremes} 
       (Dover, New York 2004).

\bibitem{jk}
      J.~Krug,
      J. Stat. Mech. P07001 (2007).

\bibitem{gw}
      G.~Wergen,
      J. Phys. A {\bf 46}, 223001 (2013).

\bibitem{ekbhs} 
       J.~F.~Eichner, E.~Koscielny-Bunde, A.~Bunde, S.~Havlin, H.~J.~Schellnhuber,
       Phys. Rev. E {\bf 68}, 046133 (2003).

\bibitem{rp}
       S.~Redner and M.~R.~Petersen
       Phys. Rev. E {\bf 74}, 061114 (2006) 

\bibitem{nmt}
      W.~I.~Newman, B.~D.~Malamud, and D.~L.~Turcotte,
      Phys. Rev. E {\bf 82}, 066111 (2010).

\bibitem{wzxw}
       Q.~H.~Wen, X.~Zhang, Y.~Xu, and B.~Wang, 
       Geophys. Res. Lett/ {\bf 40}, 1171 (2013).

\bibitem{vzm}
      Richard M. Vogel, Antigoni Zafirakou-Koulouris, and Nicholas C. Matalas
      Water Resources Research {\bf 37}, 1723 (2001) 

\bibitem{ekm}
	P.~Embrechts, G.~Kl\"{u}ppelberg and T.~Mikosch,
	{\it Modelling extremal events for insurance and finance} (Springer-Verlag, Berlin, 1997)

\bibitem{syn}
      S.~Y.~Novak,  
      {\it Extreme value methods with applications to finance}
      (Chapman \& Hall/CRC Press, London, 2011). 

\bibitem{bp}
      J.~-P.~Bouchaud and M.~Potters, 
      {\em Theory of Financial Risk and Derivative Pricing}
      (Cambridge University Press, Cambridge 2003).

\bibitem{fwk}
      J.~Franke, G.~Wergen, and J.~Krug,
      J. Stat. Mech. P10013 (2010)

\bibitem{wbk}
      G.~Wergen, M.~Bogner, and J.~Krug,
      Phys. Rev. E {\bf 83}, 051109 (2011).

\bibitem{ss}
	S.~Sabhapandit,
	EPL {\bf 94} 20003 (2011)
\bibitem{wms}
      G.~Wergen, S.~N.~Majumdar, and G.~Schehr,  
      Phys. Rev. E {\bf 86}, 011119 (2012).

\bibitem{wvrk}
        G.~Wergen, D.~Volovik, S.~Redner, and J.~Krug, 
        Phys. Rev. Lett. {\bf 109}, 164102 (2012).  

\bibitem{ekmb}
        Y.~Edery, A.~B.~Kostinski, S.~N.~Majumdar, and B.~Berkowitz,
	Phys. Rev. Lett. {\bf 110}, 180602 (2013).

\bibitem{sr}
      S.~Redner,
      {\it A Guide to First-Passage Processes}
      (Cambridge University Press, New York, 2001).

\bibitem{bms}
      A.~J.~Bray, S.~N.~Majumdar, and G.~Schehr, 
      Adv. Phys. {\bf 62}, 225 (2013).

\bibitem{bk} 
	E.~Ben-Naim and P.~L.~Krapivsky, 
	Phys. Rev. E, accepted (2013).

\bibitem{krb}  
         P.~L.~Krapivsky, S.~Redner and E.~Ben-Naim,
         {\it  A Kinetic View of Statistical Physics}
         (Cambridge University Press, Cambridge, UK, 2010).

\bibitem{gkp} 
   R.~L.~Graham, D.~E.~Knuth, and O.~Patashnik, 
   {\it Concrete Mathematics : A Foundation for Computer Science}
   (Reading, Mass.: Addison-Wesley, 1989).

\bibitem{bhi}
       E.~Ben-Naim, M.~B.~Hastings, and D.~I.~Izraelevitz,
       J. Phys. A {\bf 40}, F1021 (2007).

\bibitem{gl}
       C.~Godr\'eche and J.~M.~Luck,
       J. Stat. Mech. P10013 (2010).

\bibitem{ft}
       R.~A.~Fisher and L.~H.~C.~Tippett, 
       Proc.~Cambridge Phil.~Soc.~{\bf 24}, 180 (1928).

\bibitem{eig1}
        E.~I.~Gumbel, 
	Ann.~Inst.~Henri Poincar\'e {\bf 5}, 115 (1935).


\bibitem{ba}     
      F.~Bashforth and J.~C.~Adams, 
      {\it Theories of Capillary Action} 
      (Cambridge University Press, London, 1883). 

\bibitem{nr}
       W.~H.~Press, S.~A.~Teukolsky, W.~T.~Vetterlin, and B.~P.~Flannery, 
      {\it Numerical Recipes: The art of Scientific Computing} 
      (Cambridge University Press, London, 2007).        
       
\bibitem{am}
       A.~J.~Michael, 
       Geophys. Res. Lett. 38, L21301 (2012).
       
\bibitem{st}
       P.~M.~Shearer and P.~B.~Stark, 
       Proc. Nat. Acad. Sci. {\bf 109}, 717 (2012).
       
\bibitem{bdj}
       E.~Ben-Naim, E.~G.~Daub, and P.~A.~Johnson, 
       Geophys. Res. Lett. 40, 3021 (2013); 
       E.~G.~Daub, E.~Ben-Naim, R.~A.~Guyer, and P.~A.~Johnson, 
       Geophys. Res. Lett. 39, L06308 (2012). 
       
\bibitem{gr} 
       B.~Gutenberg and C.~F.~Richter, 
       {\it Seismicity of the Earth and Associated Phenomena} 
       (Princeton University Press, Princeton, 1954)

\bibitem{data} The earthquake catalog is available from 
{\tt http://earthquake.usgs.gov/monitoring/anss/}.

\bibitem{bk1}
        E.~Ben-Naim and P.~L.~Krapivsky,
        J. Phys. A {\bf 43}, 495008 (2010).

\bibitem{bkr} 
     E.~Ben-Naim, P.~L.~Krapivsky, and S.~Redner,
     Phys. Rev. E {\bf 50}, 822 (1994).

\bibitem{kj}
      J.~Krug and K.~Jain,
      Physica A {\bf 358}, 1 (2005).



\end{thebibliography}
\end{document}